\documentclass[12pt]{article}
\usepackage{jheppubmod}
\usepackage{mathtools}
\usepackage[labelsep=quad]{subcaption}
\usepackage[font={small,it}]{caption}
\mathtoolsset{showonlyrefs}
\usepackage{amsmath}
\usepackage{url}
\usepackage[hyphenbreaks]{breakurl}

\usepackage{array}
\usepackage{booktabs}
\usepackage{graphicx}
\usepackage{color}
\usepackage{mathtools}
\usepackage{epstopdf}

\newcommand{\be}{\begin{equation}}
\newcommand{\ee}{\end{equation}}
\newcommand{\ba}{\begin{align}}
\newcommand{\ea}{\end{align}}
\newcommand{\ry}{(\text{ry})^{-1}}
\newcommand{\bs}{\begin{split}}
\def\sess\end{split}

\def\rcomp{\tilde{R}}
\def\Tcomp{\widetilde{T}_R}
\def\ncomp{\widetilde{n}_R}
\def\lcomp{\widetilde{\lambda}_R}
\def\teach{T_1}
\def\plist{\{\alpha\}}
\def\nlist{\{N_i\}}
\def\nconstpni{\Upsilon}
\def\nconst{{\cal N}}
\def\probdens{{\mathcal P}}
\def\cumprob{{\mathcal C}}

\preprint{}
\title{Estimating the frequency of nuclear accidents}
\author{Suvrat Raju\footnote{The views expressed in this paper are my own, and do not reflect the position of my institution.}}
\affiliation{International Centre for Theoretical Sciences \\ Tata Institute of Fundamental Research \\ TIFR Centre Building \\ Indian Institute of Science Campus \\ Bengaluru 560012}
\emailAdd{suvrat.raju@gmail.com}
\date{}
\abstract{
We used Bayesian methods to compare the predictions of  probabilistic risk assessment  --- the theoretical tool used by the nuclear industry to predict the frequency of nuclear accidents --- with empirical data. The existing record of accidents with some simplifying assumptions regarding their probability distribution is sufficient to rule out the validity of the industry's analyses at a very high confidence level. We show that this conclusion is robust against any reasonable assumed variation of safety standards over time, and across regions.  The debate on nuclear liability indicates that the industry has independently arrived at this conclusion. We pay special attention to the Indian situation, where we show that the existing operating experience provides insufficient data to make any reliable
claims about the safety of future reactors. We briefly discuss some policy implications. \\
}
\keywords{Probabilistic Risk Assessment, Bayesian Analysis, Nuclear Safety}
\begin{document}
\maketitle

\section{Introduction}
This paper was motivated by the recent public debates on the safety of nuclear reactors in India. Although, this an old international debate, it has occupied a rather prominent public position in India after the passage of the Indo-US nuclear deal. What is significant is that the debate has not been confined to the technical community, but has seen the active participation of peoples' movements and civil society groups. This work attempts to parse some of the claims made by the Indian nuclear establishment and international nuclear vendors in this debate. Of course, while we have used the Indian discussion to provide motivation and also specialized some of our conclusions to India, the central conclusion that we draw regarding the conflict between the results of probabilistic risk assessment and empirical data are valid more broadly.

In this paper, we analyze a framework called ``probabilistic risk assessment'' (PRA)  that is used by the nuclear industry to calculate the expected frequency of nuclear accidents. As we review in section \ref{secprareview} below, the stated results for this expected frequency are often extremely low. Although it likely as we describe in \ref{secliability} that the industry and policy-makers are privately aware of the unreliability of these figures, they are nevertheless used quite commonly in policy-debates. Thus, for 
example, the previous chairperson of the Indian Atomic Energy Commission declared that the chance of a nuclear accident in India was ``1-in-infinity'' \cite{banerjeeinfinity}. 

Similar statements can be found in the scholarly literature.  
For example, in a review of the 1000 MW VVER reactors that were installed in the South Indian city of Kudankulam,  officials from the Nuclear Power Corporation of India (NPCIL)  pointed out that the Core-Damage Frequency (CDF) of these reactors --- the rate at which the reactor is expected to suffer accidents that damage its core --- was just $10^{-7}$ per reactor-year (ry) \cite{agrawal2006vvers}.

The other two multinational companies that are in line to construct reactors in India --- the French company, Areva, whose European Pressurised Reactors (EPRs) have been selected for Jaitapur (Maharashtra) and the the American company Westinghouse, whose AP1000 reactors have been selected for Mithi Virdi (Gujarat) --- have also put forward similar figures. For example, Areva claims that the CDF of the EPR  is 
$7.08 \times 10^{-7} \ry$. (See p. 2 of  \cite{ukpsaresults}.)
Westinghouse has estimated the CDF of the AP1000 to be $5.09 \times 10^{-7} \ry$. (See Table 5-1 and Table 8-2 of  \cite{ukhseap1000}.)

Prima facie, it is prudent to treat such low numbers skeptically, especially when they are provided with a precision of two decimal places. A nuclear reactor is a very complex system, and our understanding of its dynamics, particularly when they are coupled to an uncertain external environment, is not advanced 
enough to permit such accurate estimates.

Nevertheless, these numbers are taken seriously by regulators, both 
in India and elsewhere. In the United States, the Nuclear Regulatory Commission  has set a goal for PRA estimates of both the CDF and the large-release frequency (LRF). The latter involves accidents where the containment fails in addition to core damage.  These are expected to be below $10^{-4} \ry$ and $10^{-6} \ry$ respectively. (See   \cite{nrcwebgoals} and Appendix D of  \cite{nureg1860}.) Similar quantitative criterion have been adopted in other countries \cite{oecdpra}. In India, licensing guidelines suggest that 
new plants should have a PRA-estimated CDF smaller than $10^{-5} \ry$  and a LRF lower than $10^{-6} \ry$.
(See Annexure-1 of  \cite{aerbconsentpra}.)

The methodology used to carry out PRAs in the nuclear industry, which we review in section \ref{secprareview}, is theoretically suspect. However, the purpose of this article is to point out that, even setting aside theoretical considerations, these extraordinarily low bounds can be ruled out 
 by using the existing empirical record of nuclear accidents.

There have already been eight core-damage accidents in a little more than fifteen thousand reactor-years of experience as we review in section \ref{secempirical}. Hence, the observed  
frequency of accidents
is significantly higher than that predicted by the industry's PRAs.  Moreover, and this is the crucial conclusion of this paper:  had the true 
frequency of accidents been as low as the manufacturers claim, it is exceedingly
unlikely that so many accidents could have occurred. Conversely,
the historical record on accidents implies that, even under rather 
conservative assumptions, it is possible to conclude with a very high degree
of confidence that the results of the industry's PRAs are incorrect. 

For example, we show in  section \ref{frequencypra} that in a simplified model, even with favourable assumptions for the industry,  the hypothesis that the frequency of core-damage accidents is smaller than or equal to $10^{-7}\ry$  can be ruled out with a confidence of $1.0-4.0 \times 10^{-24}$. (Said differently, within this model, we can conclude that the core-damage frequency of nuclear reactors is higher than $10^{-7} \ry$ with a confidence of \mbox{99.9999999999999999999996}\%!)   The precise value of $4.0 \times 10^{-24}$ is specific to this model but our broader and robust conclusion is that the accident-frequencies suggested by PRA can be ruled out, even with the limited empirical data on accidents, at an extremely high confidence level.  We perform a sensitivity analysis in section \ref{secrobust} and show that even by allowing exponential increases in safety standards over time, or significant variations across regions, it is virtually impossible to reconcile the empirical data with the PRA-frequencies. In section \ref{secindia}, we turn briefly to the specific case of India. Here we make the elementary
statistical point that the existing experience of Indian reactor operation is  insufficient to derive any strong conclusions about expected accident frequencies in the future. In section \ref{secliability}, we briefly discuss the debate on liability because it shows that the nuclear industry is itself aware that the quantitative results of PRA cannot be taken seriously for commercial questions like liability insurance.
We conclude
in section \ref{secconclude}.

\section{Summary and Methodology \label{secsummary}}
In this section, we provide a short summary of our objectives, methodology and results.

Our central objective in this paper is to use the empirical data to estimate the probability for the hypothesis that the true frequency of accidents is indeed as low as that claimed by PRA. We denote this probability by $\cumprob(\lambda_{\text{pra}}, n_{\text{obs}})$ where $\lambda_{\text{pra}}$ is a frequency predicted by PRA and $n_{\text{obs}}$ is the observed number of accidents of a certain type. Our central result is that
\be
\label{centralclaim}
\cumprob(\lambda_{\text{pra}}, n_{\text{obs}}) = \epsilon \ll 1.
\ee

The precise value of $\epsilon$ depends on many factors: the precise PRA being considered, which leads to variations in $\lambda_{\text{PRA}}$, the way in which accidents are counted which leads to variations in  $n_{\text{obs}}$ and, of course, the assumptions that go into modeling nuclear accidents and the variation of safety-standards over time and across regions. 

But the point that we want to emphasize in this paper is the following:  $\epsilon$ is extraordinarily small, and this fact is robust against virtually any variation of the considerations above. The calculations presented in this paper are all calculations of $\epsilon$ under different assumptions and they all have the feature that $\epsilon$ is negligible.

Note that we are not interested in using the empirical data to compute the true frequency of accidents. Some efforts have been made in this direction \cite{sornette2013exploring,ha2014calculating} but the basic difficulty is that the empirical data is too limited to allow for a reliable computation of this frequency. The importance of focusing on $\epsilon$ is that these intermediate uncertainties do not affect the robust nature of \eqref{centralclaim}.

Second note that to compute $\epsilon$, we are forced to use Bayesian analysis. While a purely frequentist approach might demonstrate that the empirically observed frequency of accidents does not agree with the frequency predicted by PRA this does not, by itself, tell us the confidence level with which we can rule out the correctness of PRA. In fact the question of estimating the confidence-level cannot be posed within the frequentist approach at all. This confidence-level is given by $1 - \epsilon$, and requires Bayesian methods. 

We now turn to the main body of the paper.

\section{A Brief Review of Probabilistic Risk Assessment \label{secprareview}}

The systematic use of PRA in the nuclear industry started with the Rasmussen report of 1975 \cite{rasmussen}. This  report was commissioned by the U.S. Nuclear Regulatory Commission (NRC) and met with criticism soon after its publication.  In 1977, the U.S. NRC commissioned a review of the Rasmussen report through the Lewis Committee,  and following this critical review-report, released a statement in 1978 stating that ``the Commission does not regard as reliable the Reactor Safety Study's numerical estimates of the overall risk of reactor accidents.'' (See Pg. 533 of \cite{kok2009nuclear}.) In spite of these early objections, over the past few decades, the framework of probabilistic risk assessment has become influential in propagating the use of quantitative probabilistic techniques for questions of nuclear safety.  Moreover, as described above, the use of these techniques is not confined to qualitative safety analyses aided by quantitative probabilistic techniques, but rather the actual numerical results produced by PRA are used by regulators, and in policy debates.

The basic idea used by such studies is simple to describe: one enumerates the possible fault trees that could lead to an accident.
For each individual component in the reactor, one can estimate a frequency of failure. For a serious accident, some combination of these components has to fail simultaneously. The industry advertises a philosophy of ``defense in depth'', which reduces the overall possibility of an accident by building redundancy into a system. The low numbers above result from the fact that several systems have to fail simultaneously before the core is damaged.

For example, Keller and Modarres  \cite{keller_historical_2005}  report that in 1966, in one of the earliest such assessments, the General Electric company ``showed'' that its reactors ``had a one-in-a-million chance per year for a catastrophic failure because each of the three major subsystems would only fail once-in-one-hundred per years'' {\em (sic)}. The Rasmussen report, which followed a decade later, attempted to refine and formalize this methodology.

The theoretical problem with such estimates is obvious. Consider
the Fukushima nuclear complex, which had 13 backup diesel generators \cite{wnnfukushima}. Assigning a probability of $10^{-1}$ for the failure of each generator per year and assuming that they are independent  would lead us to the 
naive conclusion that the probability that 12 generators would fail together in any given year is about $13  \times 10^{-12} \times 0.9 \approx 10^{-11}$. 
However, the tsunami did precisely this by disabling all but one of the generators at once.  The point is that once the obvious fault trees have been eliminated and corrected, we reach a stage where
 the dominant contributions to accident-probabilities come from unlikely sequences of events that conspire to cause a failure. 

This issue occurs in any suitably complex system. However, a nuclear reactor is also an open system that is coupled
to an external environment. This makes it virtually impossible to foresee all sorts of low probability
pathways to failure. Furthermore,  our understanding of the frequency of extreme initiating events, such as tsunamis and earthquakes, is itself rather crude.
These error bars overwhelm 
the seemingly accurate predictions that result from elaborate simulations of the reactor. 

However, instead of venturing deeper into these theoretical arguments --- which are considered in greater detail elsewhere in the literature (see, for example,  \cite{diazmaurinthesis}) --- we will examine how the predictions of PRA  stand up against the extant empirical data. 

To facilitate this comparison, let us summarize the various kinds of claims that
have been made by the three multinational companies relevant to India. 
As we mentioned, Areva has estimated that the CDF of 
an EPR accounting for both internal and external hazards is $7.08 \times 10^{-7} \ry$. The LRF of the EPR is estimated to be about 11\% of its CDF: $7.69 \times 10^{-8} \ry.$ (See p. 14 of  \cite{ukpsaresults} and p. 222 of \cite{ukpsaresultslevel2}.)

Similarly, Westinghouse has estimated that the CDF of the AP1000 is 
$5.09 \times 10^{-7} \ry$.
The LRF of this reactor
is estimated to be $5.94 \times 10^{-8} \ry$. (See Table 8-2 of  \cite{ukhseap1000}.) 
This includes accidents due to ``external hazards'' including ``external flooding, extreme winds, seismic, and transportation accidents''. In fact Westinghouse concluded that ``conservative bounding assessments show that core damage risk from events listed above is small compared to the core damage risk from at-power and shutdown events.'' (See p. 4-37 of \cite{ukhseap1000}.)

Although 
the NPCIL has stated that the CDF of the Kudankulam VVER reactors is $10^{-7} \ry$, we were unable to locate
the details of the PRA that led to this conclusion. So, we will take this figure as is, and consider a LRF of $10^{-8} \ry$ using the common estimate that the LRF ``is generally about ten times less than CDF'' \cite{wnacdflrf}. All these claims are summarized in Table \ref{tablepra}.
\begin{table}[!h]
\begin{center}
\resizebox{\textwidth}{!}{
\begin{tabular}{ccc}
\toprule {\bf Reactor (Manufacturer)} & {\bf Core-Damage Frequency}  & {\bf Large-Release Frequency} \\ \midrule
EPR (Areva) &  $7.08 \times 10^{-7} \ry $ 
&  $7.69 \times 10^{-8} \ry$  
\\ 
AP1000 (Westinghouse)&$5.09 \times 10^{-7} \ry $ 
& $5.94 \times 10^{-8} \ry$ 
\\ 
VVER (Rosatom) &  $10^{-7} \ry$ 
& $10^{-8} \ry$  
\\ \bottomrule
\end{tabular}}
\caption{Predictions of Probabilistic Risk Assessment \label{tablepra}}
\end{center}
\end{table}

\section{Review of the Empirical Experience \label{secempirical}}
We now review the historical record on nuclear accidents. The industry, as a whole, had gathered about $T_{\text{obs}} = 15247$ reactor-years of operating experience by the end of 2012 according
to the latest data put out by the International Atomic Energy Agency (IAEA). (See Table 4 of \cite{npriaea2013}.)
In this time, there have been
several core-damage accidents. 

Surprisingly, the IAEA does not maintain a comprehensive historical record
of core-damage accidents. Cochran and McKinzie have compiled a very useful
list of 25 such instances from various sources \cite{cochranmckinziw}. In our analysis, we
will only consider accidents that occurred at commercial reactors, thus excluding
accidents at experimental facilities like Enrico Fermi Unit-1 (1966) or Lucens (1969). Even this enumeration is somewhat subjective, but a conservative approach,
keeping only accidents that involved a significant meltdown of fuel leads
to  the list in Table \ref{nuclearacc}. In each case, we have also provided references to more detailed descriptions of these accidents.
\begin{table}
\resizebox{\textwidth}{!}{
\begin{tabular}{p{1.4in}p{0.7in}p{0.5in}p{1in}p{1.9in}}
\toprule
{\bf Reactor} & {\bf Country} & {\bf Year} & {\bf Reference} & {\bf Note} \\
\midrule 
Saint-Laurent A-1 & France & 1969 & p. 35 of \cite{selectednuclearacc} & Meltdown of 50 kg of fuel.  
\\[10pt] 
Three Mile Island Unit 2 & USA & 1977 & \cite{kemenycommissiontrunc} & Severe accident; radiation release. 
\\[10pt] 
Saint-Laurent A-2 & France &1980 & Table 12 of \cite{inesdescription} and \cite{cochranmckinziw} & Meltdown of one channel of fuel. 
\\[10pt] 
Chernobyl Unit 4 & Ukraine & 1986 & \cite{_chernobyl_1992} & Severe accident; large radiation release. 
\\[10pt] 
Greifswald Unit 5 & Germany (GDR) & 1989 & \cite{wnanuclearsafetygreif,wnadecommgreif,widegreif,cochranmckinziw} & Partial core meltdown soon after commissioning.
\\[10pt] 
Fukushima Daiichi Units 1,2,3 & Japan & 2011 & \cite{fukushimaindependentreport} & Severe accident; large radiation release. \\[10pt] 
\bottomrule
\end{tabular}}
\caption{List of core-damage accidents in commercial reactors \label{nuclearacc}}
\end{table}

Table \ref{nuclearacc} enumerates accidents at $n^{\text{cd}}_{\text{obs}} = 8$ reactors. 
Of these 8 accidents, $n^{\text{lr}}_{\text{obs}} = 5$ accidents  led to the release of large amounts of radioactive substances into the environment. This list comprises the accidents at Three Mile Island, Chernobyl and the three at Fukushima. Since we have PRA results for
both the CDF and the LRF, we can compare these separately to the historical record.

The reason for counting the three accidents at Fukushima separately in the list above is that we are interested in the rate of accidents per reactor-year of operation, and all these three reactors contribute separately
to the ``total operating experience'' that appears in the denominator. 

We comment more on this issue in section \ref{secconclude}. However, to demonstrate the robust nature of our conclusions, 
we will present a parallel analysis where the accidents at Fukushima are counted together. By this (incorrect) counting, there have been $n^{\text{cd}}_{\text{low}} = 6$ core-damage accidents and $n^{\text{lr}}_{\text{low}}=3$ accidents with a large release of radioactivity. 

\section{Simplified Bayesian Analysis of  Empirical Frequencies and PRA  \label{frequencypra}}
Table \ref{nuclearacc} leads to the following observed frequency of core-damage and large-release accidents.
\be
\label{observedfreq}
\begin{split}
&\nu^{\text{cd}}_{\text{obs}} = {n^{\text{cd}}_{\text{obs}} \over T_{\text{obs}}} \approx {1 \over 1906} \ry \approx 5.2 \times 10^{-4} \ry  , \\
&\nu^{\text{lr}}_{\text{obs}} = {n^{\text{lr}}_{\text{obs}} \over T_{\text{obs}}} \approx {1 \over 3049} \ry \approx  3.3 \times 10^{-4} \ry  .
\end{split}
\ee
Counting the Fukushima accidents as a single accident, we find
\be
\begin{split}
&\nu^{\text{cd}}_{\text{low}} = {n^{\text{cd}}_{\text{low}} \over T_{\text{obs}}} \approx {1 \over 2541} \ry \approx 3.9 \times 10^{-4} \ry  , \\
&\nu^{\text{lr}}_{\text{low}} = {n^{\text{lr}}_{\text{low}} \over T_{\text{obs}}} \approx {1 \over 5082} \ry  \approx 2.0 \times 10^{-4} \ry  .
\end{split}
\ee

It is clear that these empirically observed frequencies are far higher than
the predictions of the manufacturers' PRAs. However, we can ask a more detailed question: given the observed rate of accidents, what is the probability that the results of PRA are close to, or smaller than, the ``true frequency'' of accidents.

Bayesian techniques are ideally suited to answer this question, and we return to this issue again below. To answer the question above,

we need
to make a few simplifying assumptions about the probability distribution of nuclear accidents. As a first approximation, we start by assuming that nuclear accidents are independent
events, and that in every small time interval $d t$, each reactor has
 a small and constant probability 
\be
\label{centralassumption}
d p = \lambda \,  d t,
\ee
of suffering an accident.  The use of this approximation is {\em not meant} to suggest that it is truly the case that accident frequencies have not changed over time. Rather, as we show in section \ref{secrobust} our results are very {\em robust} against almost any reasonable assumed changes in safety standards over time, and across regions.

Therefore, given this robustness of our central results, we present this simplified model first since it is amenable to simple analysis and already captures the central points that we wish to make. The reader who is interested in a more sophisticated analysis should consult the next section.

Say that we have $m$ reactors functioning simultaneously which we observe for a time period $\teach = N d t$.  The probability for $n$ of these to have undergone accidents is 
\be
\label{derivprobdistrib}
\begin{split}
p_{\lambda}(n) &=  \begin{pmatrix} m \\ n \end{pmatrix} \big(1 - \lambda d t)^{N (m-n)}  \Big(1-\big(1 - \lambda d t \big)^N \Big)^n \\ &= \begin{pmatrix} m \\ n \end{pmatrix} \big(1 - {\lambda \teach \over N } \big)^{N(m-n)} \Big(1-\big(1 - {\lambda \teach \over N }\big)^N \Big)^n \\
&\underset{N \rightarrow \infty}{\longrightarrow} {\Gamma(m+1) \over \Gamma(m-n+1) \Gamma(n+1)} e^{-\lambda \teach m (1 - {n \over m})} \big(1 - e^{-\lambda \teach} \big)^n,\\ 
\end{split}
\ee
where, in the last line, we have taken the continuous limit: $N \rightarrow \infty$, with $\teach$ finite. We remind the reader that $\Gamma(n+1) = n!$ is the standard Gamma function.
We now consider the case where $\lambda \teach \ll 1$, which implies that the chance of any individual reactor undergoing an accident is small, and $n \ll m$, which states  that only a small fraction of all reactors undergo an accident. In this limit, the total operating experience gathered becomes $T = m \teach$, and we see that the distribution simplifies to
\be
\label{poisson}
p_{\lambda}(n) =  
{1 \over \Gamma(n+1)} (\lambda T)^{n} e^{-\lambda T}.
\ee
 This is just a Poisson distribution and we could even have started with this distribution,  which is
commonly used to model accidents and other {\em rare events} in various scenarios. 

It is
 now possible to solve the following Bayesian problem: start with the {\em prior} assumption
that $\lambda$ is uniformly distributed
\be
\label{priorprob}
\probdens(\lambda) = {\theta(\lambda) - \theta(\lambda - \lambda^c)\over \lambda^c}, 
\ee
where $\lambda^c$ is an irrelevant high frequency cutoff to make the distribution normalizable.
Given the observed
frequency of events above, what is the {\em posterior}
probability distribution for $\lambda$?

Let us try and explain this in simple words. Say we start with no prior bias
for the value of $\lambda$. Given that we have some number $n_{\text{obs}}$ of accidents, 
we can use this empirical data to form an estimate as to the value of $\lambda$. In fact, Bayes theorem
tells us  that we can actually calculate the probability that the true frequency of
accidents has a value between $\lambda$ and $\lambda + d \lambda$ through the formula
\be
\label{bayestheorem}
\probdens(\lambda|n_{\text{obs}}) = {\probdens(n_{\text{obs}} | \lambda) \probdens(\lambda) \over \probdens(n_{\text{obs}})}.
\ee
This formula is just making precise that intuition that the empirical evidence is giving us
some information about the value of $\lambda$ in our world.

On the right hand side, $\probdens(n_{\text{obs}} | \lambda)=p_{\lambda}(n_{\text{obs}})$, which is given by the Poisson distribution \eqref{poisson}. $\probdens(\lambda)$ is our flat prior probability distribution in \eqref{priorprob}. To fix $\probdens(n_{\text{obs}})$, which 
is a $\lambda$ independent constant, we can simply demand that
$\int_0^{\infty} \probdens(\lambda | n_{\text{obs}}) d \lambda  = 1$. 

\begin{figure}[!h]
\begin{center}
\includegraphics[width=0.6\textwidth]{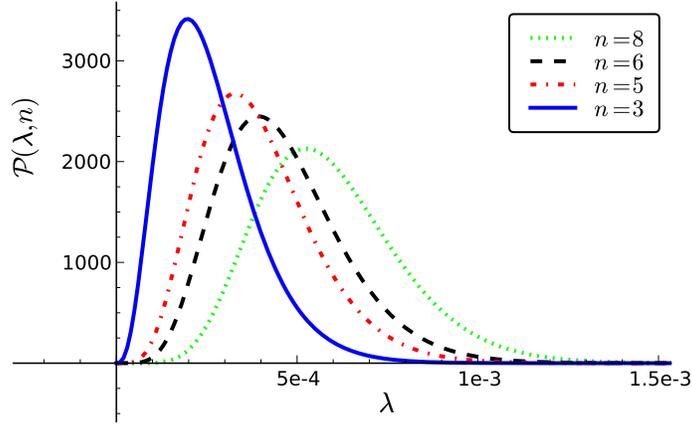}
\caption{Posterior probability distribution for the parameter $\lambda$ \label{plambda}}
\end{center}
\end{figure}

After fixing this constant we find 
that the posterior
probability distribution for $\lambda$ is given by
\be
\label{posterior}
\probdens(\lambda | n_{\text{obs}}) = {1 \over \Gamma(n_{\text{obs}} + 1) }T_{\text{obs}} (\lambda
T_{\text{obs}})^{n_{\text{obs}}} e^{-\lambda T_{\text{obs}}},
\ee
where we have neglected terms of $\text{O}\big(e^{-\lambda_c T_{\text{obs}}}\big)$, assuming $\lambda_c$ is taken to be large enough. 
This function is plotted in figure
\ref{plambda} for the values of $n_{\text{obs}}$ that are relevant to both
the large-release  and the core-damage frequency.

We pause to mention a somewhat subtle point.  Since the observed number of 
 events is small, $n_{\text{obs}} \sim \text{O}\left(1\right)$, the curves in figure \ref{plambda} have an appreciable width. This indicates the difficulty with using a {\em frequentist} approach to estimating the true frequency of accidents using empirical data. However, as we emphasized in section \ref{secsummary} if we are interested in estimating the parameter $\epsilon$ in \eqref{centralclaim} instead, then Bayesian methods yield a robust statement. We now turn to this calculation.

\begin{figure}[!h]
 \begin{center}
\includegraphics[width=0.6\textwidth]{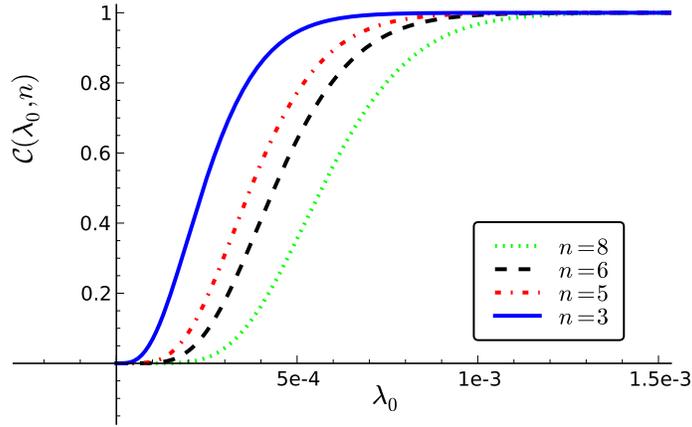}
\caption{Probability for the hypothesis $\lambda < \lambda_0$ \label{cumprob}}
\end{center}
\end{figure}

To estimate $\epsilon$,  we use \eqref{posterior} to calculate the probability that
the probability that $\lambda$ is smaller than any given
$\lambda_0$. This function is given by
\be
\label{cumlambda}
{\cumprob}(\lambda_0, n_{\text{obs}}) = \int_0^{\lambda_0} \probdens(\lambda|n_{\text{obs}}) \, d \lambda= 1 -
{\Gamma(1 + n_{\text{obs}}, T_{\text{obs}} \lambda_0) \over \Gamma(1 +
  n_{\text{obs}})}.
\ee
where $\Gamma(k,z)$ is the incomplete gamma function.  The function 
$\cumprob(\lambda_0, n)$ is shown in figure \ref{cumprob} for all 
the relevant values of $n$.

Since the probability that the true frequency is smaller than
the various results of PRA is so close to zero, it cannot be read off the graph. However, we can use the series expansion
\be
\label{incompgammafunc}
\Gamma(k,z) = \Gamma(k) - {x^k \over k} + {x^{k+1} \over k+1} + \text{O}\left(x^{k+2}\right).
\ee
This leads to the numerical figures given in Table \ref{praright}.
The phrase ``PRA ... is right'' is shorthand for the hypothesis that the true frequency of accidents
is lower than or equal to the frequency predicted by PRA . Therefore, the table lists the  values of the function $\cumprob(\lambda_0, n_{\text{obs}})$ from \eqref{cumlambda} with $\lambda_0$ set to the PRA-predicted frequency and $n_{\text{obs}}$ set to the observed number of accidents, which varies depending
on whether the Fukushima accidents are counted together or separately. Note that the values for the probabilities in this table correspond precisely to the parameter $\epsilon$ in \eqref{centralclaim}.
\begin{table}[!h]
 \resizebox{\textwidth}{!}{
\begin{tabular}{ccccc}
\toprule
{\bf Reactor}& \multicolumn{2}{c}{\bf Probability PRA CDF is right}  & \multicolumn{2}{c}{\bf Probability PRA LRF is right}  \\ \cmidrule(r){2-3}
\cmidrule(l){4-5}
&{\bf Fukushima separate} & {\bf Fukushima together} & {\bf Fukushima separate} & {\bf Fukushima together} \\ \midrule
 {Kudankulam}&$1 \times10^{-31}$&$4 \times 10^{-24}$&$ 2 \times 10^{-26}$&$2 \times 10^{-17}$ \\
 {EPR}& $5 \times 10^{-24}$ & $3  \times 10^{-18}$ & $4 \times 10^{-21}$ & $8 \times 10^{-14}$ \\  {AP1000}&$3 \times 10^{-25}$&$3 \times 10^{-19}$&$8 \times 10^{-22}$& $3 \times 10^{-14}$\\ \bottomrule
\end{tabular}}
\caption{Comparing PRA results with Bayesian estimates from historical observations (Simplified Model) \label{praright}}
\end{table}

One observes immediately that, given the empirical data, the probability
that the industry's PRA-based conclusions are right is astronomically small. As we stated 
in the introduction, this implies that with almost perfect certainty
we can conclude that the true frequency of accidents is much larger than
the figures advertised by the manufacturers. 

We emphasize that the figures in Table \ref{praright} should not be used  as precise numerical bounds on the validity of the PRA results. The precise values of $\epsilon$ listed there suffers from several uncertainties that we have already mentioned. As we show in the next section upon consideration of a more sophisticated model, the numerical values of these probabilities change but the robust statement is that they always remain extremely small.

In words, the results 
of Table \ref{praright} can be stated by means of 
the following straightforward conclusion: the historical data on nuclear accidents provides overwhelming evidence that the methodology of probabilistic risk assessment is seriously flawed.  A corollary is that  the observed frequency of  accidents contradicts the industry's claim that the probability of an accident is negligible.

\section{Robustness of the Simplified Model \label{secrobust}}
In this section, we model possible improvements in safety standards over time, and variations in accident frequencies across regions. We show that the central results of the simplified Bayesian model above are very robust against any reasonable assumed variations of this kind. However, as the reader will note the analysis of this section is mathematically more involved and therefore the reader who is willing to accept this conclusion may skip this section on a first reading.

\subsection{Modelling Improvements in Safety over Time}
To model possible improvements in safety over time, we now relax the assumption made in \eqref{centralassumption} and allow the probability of an accident to vary with time. We first discuss a general framework to model this possibility and then discuss a concrete model.

\subsubsection{Framework for Time-Variations of Safety}
The probability that an accident occurs between time $t$ and $t + dt$ is given by
\be
d p(t) = \lambda_{\plist}(t) d t,
\ee
where the subscript $\plist$ indicates a set of parameters that control the variation of $\lambda$ in time. 

Now, consider a time interval of length $T$, which we divide into $N$ equal parts set off by $0 < t_1 < t_2 \ldots t_{N-1} < T$ with $t_{i+1} - t_{i} = d t$. To specify the pattern of accidents note now that a single number is not enough, but instead we require a {\em function} $N(i)$,  which tells us whether an accidents occurred in the interval $[t_i, t_{i+1}]$.  This is defined through
\be
N(i) = \left\{\begin{array}{ll}1,&\text{if~an~accident~occurs in~the~interval}~[t_i, t_{i+1}] \\ 0,&\text{otherwise}\end{array}\right.
\ee
Clearly, the probability for any such pattern of accidents is given by
\be
p(\nlist|\plist) = \nconstpni_{\plist} \prod_{N_i = 1} \lambda_{\plist}(t_i) d t \prod_{N_i = 0} (1 - \lambda_{\plist}(t_i)),
\ee
where $\nconstpni_{\plist}$ is a normalization factor that we discuss below. Note that this probability is infinitesimal for any given pattern $N(i)$, and this is not surprising since in the continuum limit, we need to do a {\em path integral} over all possible patterns of accidents. However, as we see below for our purpose of constructing the posterior probability distribution, this constant will not be important.

In the continuum limit, if $n$ accidents are observed at times $t_{i_1} \ldots t_{i_n}$, then we find that
\be
\probdens\left(\nlist | \plist\right) \propto \left[\prod_{j} \lambda_{\plist}(t_{i_j}) \right]e^{-\int_0^T \lambda(t) d t}
\ee
Then, once again using Bayes theorem, we find that
\be
\probdens\left(\plist | \nlist \right) = {\probdens\left(\nlist | \plist \right) \probdens(\plist)  \over \probdens(\nlist)}
\ee
We see therefore that the unknown normalization constants drop out and that the  {\em posterior probability distribution} for the parameters upon observation of a certain pattern of accidents is given by 
\be
\probdens\left(\plist | \nlist \right) = \nconst \left[\prod_{j} \lambda_{\plist}(t_{i_j}) \right]e^{-\int_0^T \lambda(t) d t} \probdens(\plist)
\ee
where $\probdens(\plist)$ is the {\em prior probability distribution} for the parameters and the normalization $\nconst$ is now a simple finite quantity that is simply fixed by 
\be
\int \probdens\left(\plist | \nlist \right) d \plist  = 1
\ee

\subsubsection{A Concrete Model}
We now turn to a concrete model that shows how the framework above may be utilized. We assume a variation of probability with time as 
\be
\label{varylambda}
\lambda(t) = \lambda_i e^{-\gamma t}.
\ee
Here $t$ is the total operating experience accumulated by the industry. We assume that $\gamma > 0$ and this models the frequency of accidents as starting with $\lambda_i$ and decreasing exponentially with time as the industry gains additional experience.  
This model of an exponential increase in safety standards constitutes a very favourable assumption for the industry but we will see that even allowing for this, our central conclusions are unchanged. 
Of course, the reader can easily generalize these results to more general variations of the frequency. 

Furthermore, we assume a flat prior distribution of the form \eqref{priorprob}  for both $\lambda_i$ and $\gamma$. 
\be
\label{flatpriorvary}
\begin{split}
\probdens(\lambda_i) &= {\theta(\lambda_i) - \theta(\lambda_i - \lambda_i^c)\over \lambda_i^c} \\
\probdens(\gamma) &= {\theta(\gamma) - \theta(\gamma - \gamma^c)\over \gamma^c},
\end{split}
\ee
where $\lambda_i^c$ and $\gamma^c$ are cutoffs.  As we described above, the cutoff on $\lambda_i$ is irrelevant, but in this exponential model we have to be somewhat careful about the cutoff $\gamma^c$. This is because if we assume a prior that is flat over very large ranges of $\gamma$, then this allows for a fat-tail in the posterior distribution that represents the scenario where the accident frequencies were very large in the past but have improved rapidly very recently. We will make a specific numerical choice of cutoff below, although we postpone the question for the moment. 

We should emphasize two subtleties.  First, while the cutoff, $\gamma^c$ is clearly physically important and changes the numerical values of probabilities, it it not strictly required for convergence.  Second, while the flat priors in \eqref{flatpriorvary} reflect our ignorance about these parameters this necessarily involves a choice of basis.  Note, for example, that assuming a flat prior for the initial frequency $\lambda_i$ is different from assuming a flat prior for the current frequency $\lambda(T)$.

Now consider the case where $n$ accidents have been observed at times $t_{i_1}, \ldots t_{i_n}$ in a total operating time $T$.  Define $\tau = \sum_j t_{i_j}$. Then it is clear from the analysis above that the posterior probability distribution for $\lambda_i, \gamma$ is
\be
\label{postligamma}
\probdens\left( \lambda_i, \gamma | \nlist \right) = \nconst \lambda_i^n e^{-\gamma  \tau -{\lambda_i \over \gamma}\big(1 - e^{-\gamma T} \big)}
\ee

We can determine the marginal probability distributions for both $\lambda_i$ and $\gamma$ by integrating over the other variable. In particular, the distribution for $\gamma$ can be obtained by doing the easy integral over $\lambda_i$ and is given by
\be
\probdens\left(\gamma | \nlist \right) = \int_0^{\infty} d \lambda_i \probdens\left( \lambda_i, \gamma | \nlist \right) = \nconst e^{-\gamma  \tau } \Gamma (n+1) \left(\frac{\gamma}{1-e^{-\gamma T}}\right)^{n+1}
\ee

On the other hand, it does not appear possible to write the distribution for $\lambda_i$ in terms of elementary functions. However, a double-infinite series representation can be obtained as follows by expanding the exponentials.
\be
\begin{split}
\probdens\left(\lambda_i | \nlist \right) &= \int_0^{\gamma^c} d \gamma \probdens\left( \lambda_i, \gamma | \nlist \right) \\
&=  \nconst \int_0^{\gamma^c} d \gamma \sum_{m,q=0}^{\infty} \left(\frac{(-1)^q e^{-\frac{\lambda_i}{\alpha }} \lambda_i^{m+n} \alpha ^{q-m} (m T+\tau )^q}{\Gamma (m+1) \Gamma (q+1)} \right) \\
&= \nconst \sum_{m,q=0}^{\infty} \frac{(-1)^q \lambda_i ^{n+q+1} (m T+\tau )^q \Gamma \left(m-q-1,\frac{\lambda_i}{\gamma^c}\right)}{\Gamma (m+1) \Gamma (q+1)}.
\end{split}
\ee

By integrating these distributions, we also obtain the value of the normalization constant $\nconst$ through
\be
\begin{split}
1 &= \int \probdens\left(\gamma | \nlist \right) d \gamma = \int \probdens\left(\lambda_i | \nlist \right) d \lambda_i  \\ &= \nconst \sum_{q=0}^{\infty}\frac{\Gamma (n+q+1) (q T+\tau )^{-n-2} (\Gamma (n+2)-\Gamma (n+2,\gamma^c  (q T+\tau )))}{\Gamma (q+1)}
\end{split}
\ee
which yields an expression for $\nconst$ as the inverse of an infinite sum.

Another interesting quantity is the posterior probability distribution for the ``current accident frequency'' $\lambda_T = \lambda_i e^{-\gamma T}$. We can change variables in \eqref{postligamma}, after including a Jacobian factor, to obtain
\be
\label{postltgamma}
\probdens\left( \lambda_T, \gamma | \nlist \right) = \nconst \lambda_T^n e^{-\gamma  \tau +\frac{\lambda_T \left(1-e^{\gamma  T}\right)}{\gamma }+\gamma  (n+1) T}
\ee

As usual the probability that $\lambda_i$ is smaller than a given value, $\lambda_0$, or the probability that $\gamma$ is smaller than some $\gamma_0$ is given by integrating the probability distributions above.
\be
\begin{split}
&\cumprob_{\gamma}(\gamma_0 , \nlist) = \int_0^{\gamma_0} \probdens\left(\gamma | \nlist \right) d \gamma \\
&\cumprob_{i}(\lambda_0  , \nlist) = \int_0^{\lambda_0} \probdens\left(\lambda_i | \nlist \right) d \lambda_i  \\
&\cumprob_{T}(\lambda_T^0 , \nlist) = \int_0^{\lambda_T^0} d \lambda_T \int_0^{\gamma^c} d \gamma \probdens\left( \lambda_T, \gamma | \nlist \right)
\end{split}
\ee
While it is possible to write these expressions in terms of an infinite series of elementary functions, these forms are too complicated to be useful. It is, of course, possible to evaluate these expressions numerically at any given point as we do below. 

We now turn to the numerical values of these expressions at the empirically relevant points. Since the expressions above depend on the specific times at which the accidents occurred, we need some additional information: the reactor-years of operating experience that the industry had accumulated at the time of the accidents listed in Table \ref{nuclearacc}. Using Table 7 of \cite{npriaea2013}, we can estimate these figures for the years in which the accidents occurred, by simply adding the number of operating reactors, as given in that table. This estimate is more than sufficient for our purposes and results in the Table \ref{tableopexp}.
\begin{table}[!h]
\begin{center}
\begin{tabular}{llc}
\toprule
 {\bf Accident} & {\bf Year} & {\bf Operating Experience} (ry) \\
\midrule
Saint-Laurent A-1 & 1969 & 391 \\
Three Mile Island & 1977 & 1406 \\
Saint-Laurent A-2 & 1980 & 2048 \\
Chernobyl Unit 4 & 1984 & 3150 \\
Greifswald Unit 5 & 1989 & 5061 \\
Fukushima Units 1,2,3 & 2011 & 14572 \\
\bottomrule
\end{tabular}
\end{center}
\caption{Operating Experience (Reactor-Years) Accumulated by the Industry at the Time of Each Accident \label{tableopexp}}
\end{table}

Finally, to obtain numerical values we also need to choose a value for the cutoff on the rate of improvement. We choose
\be
\gamma^c = {\ln(50) \over T_{\text{obs}}},
\ee
which indicates our prior assumption that safety standards in the industry have improved by, at most, a factor of $50$ from early commercial nuclear reactors. Of course, the reader can consider other values of the cutoff and as we mentioned earlier it is even possible to remove the cutoff altogether.

\begin{figure}[!h]
\begin{center}
\begin{subfigure}[t]{0.4\textwidth}
\includegraphics[height=0.3\textheight]{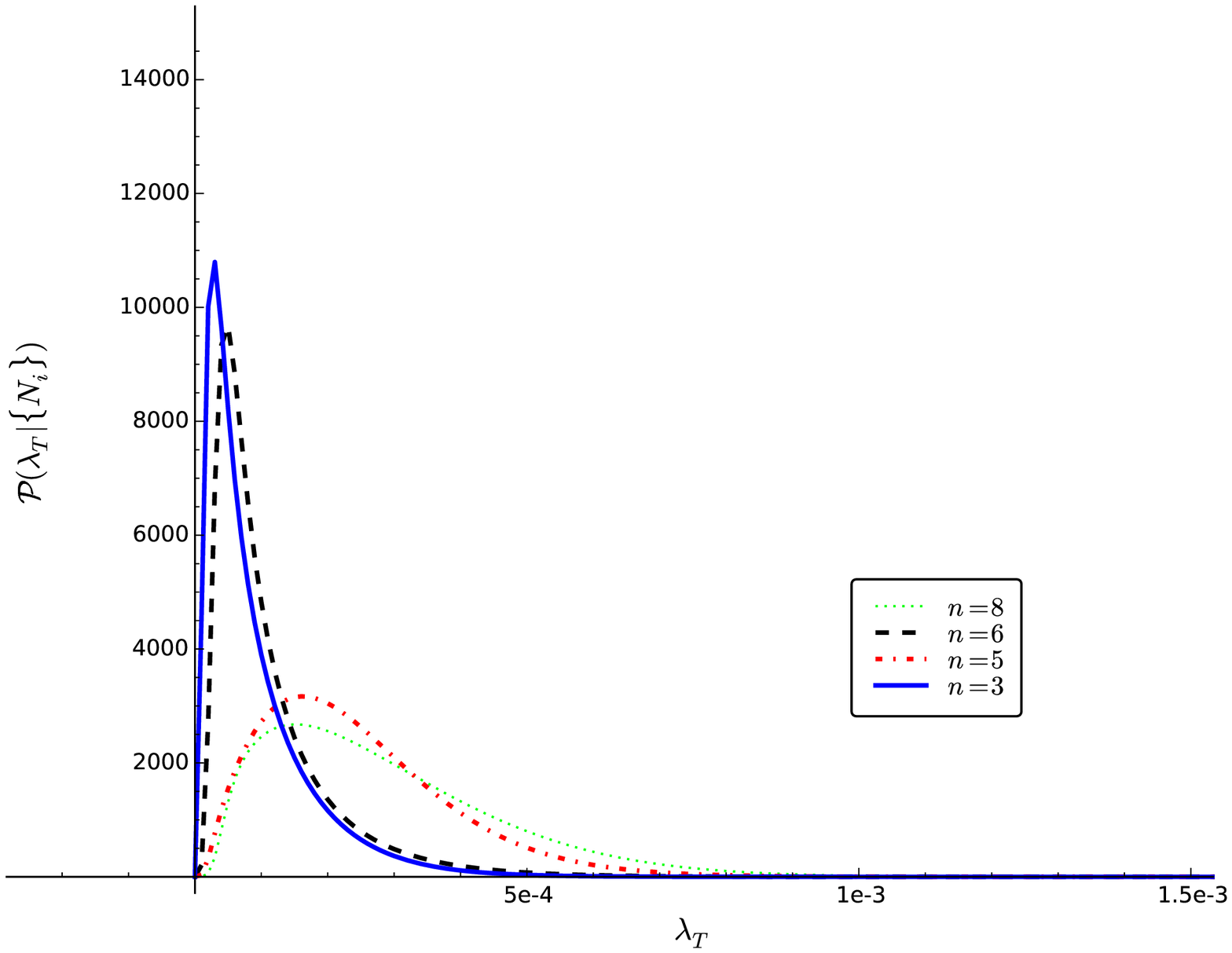}
\caption{Current Accident Frequency}
\end{subfigure}
\qquad \qquad \qquad
\begin{subfigure}[t]{0.4\textwidth}
\includegraphics[height=0.3\textheight]{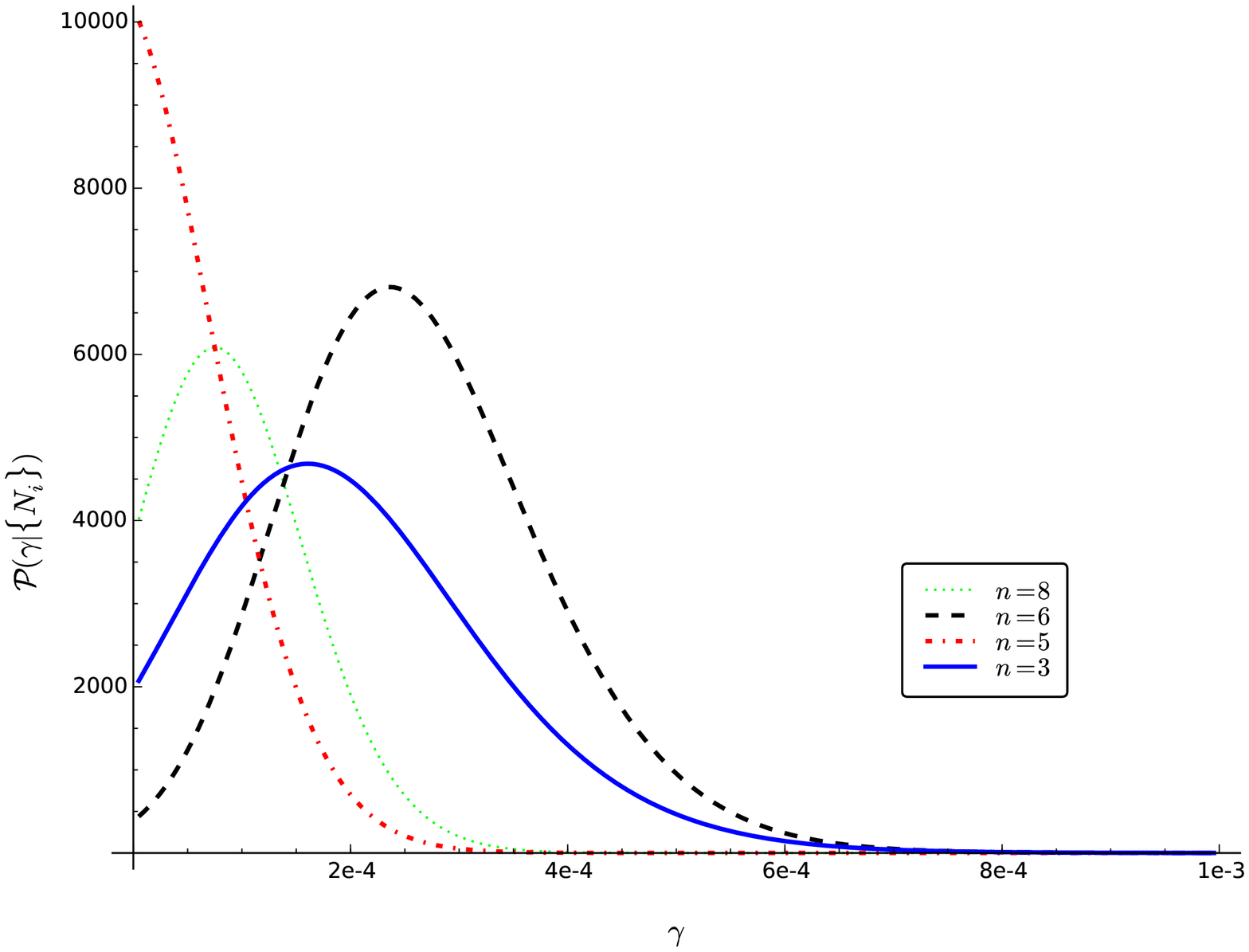}
\caption{Rate of Improvement}
\end{subfigure}
\caption{Posterior Probability Distributions for the two parameters of the model. \label{figpostvary} }
\end{center}
\end{figure}

With these numerical values, it is possible to plot the various probability distribution functions given above. The posterior probability distributions for $\lambda_T$ and $\gamma$ are plotted in Figure \ref{figpostvary}. The alert reader may note that the probability distribution for $n=5$ is peaked to the right of the distribution for $n=6$. This is because the case with $n=5$ represents the analysis for large-release accidents, where the Fukushima accidents are counted separately. This tends to disfavour large values of $\gamma$, since in this counting, $3$ out of $5$ accidents happened at late times. By thus disfavouring rapid recent improvements in safety, this particular case tends to disfavour low values of $\lambda_T$, and this is why the distribution with $n=6$ is peaked to the right of the case with $n=5$ even though the observed number of accidents is larger in this case.

\begin{figure}[!h]
\begin{center}
\includegraphics[width=0.6\textwidth]{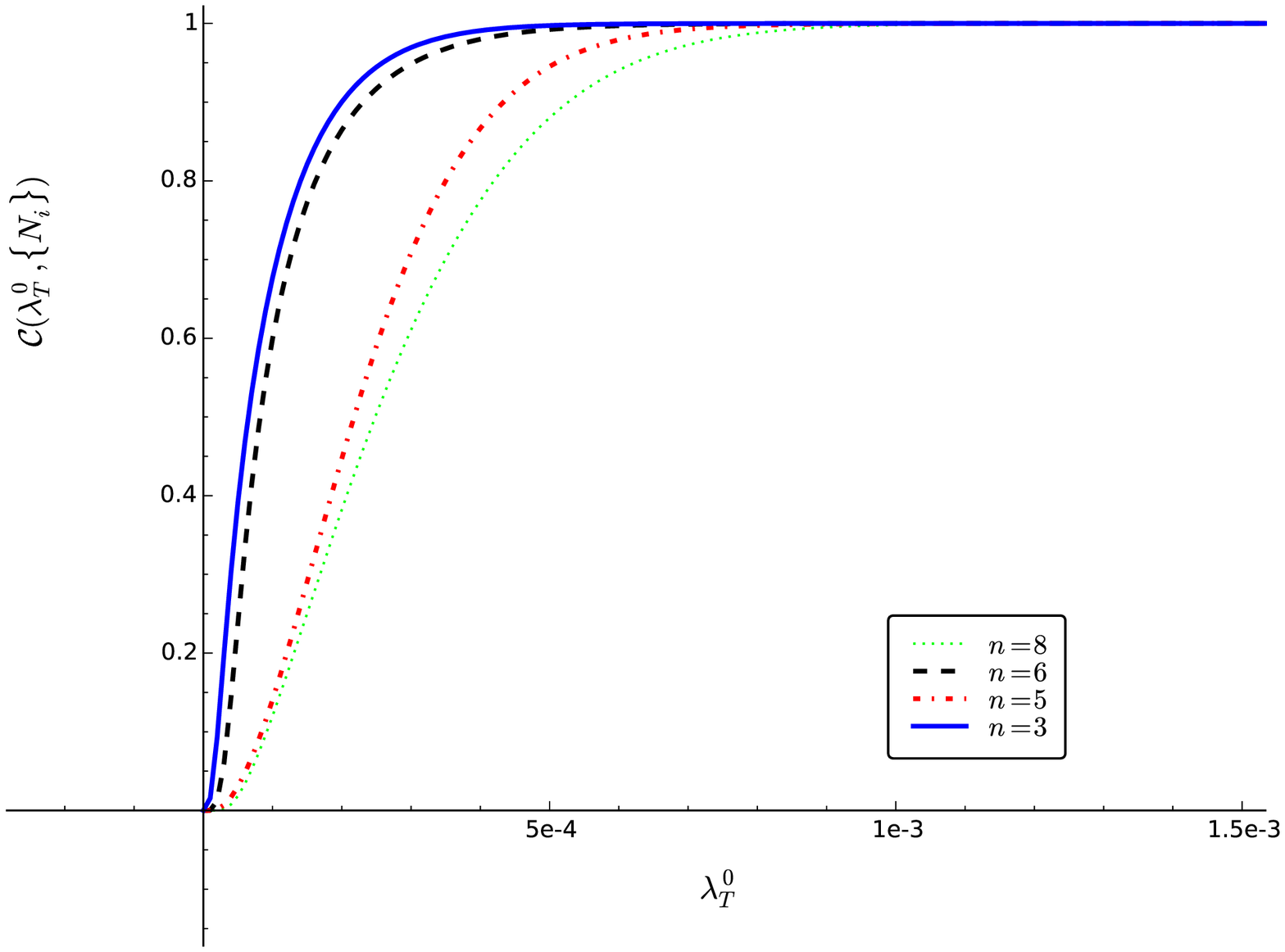}
\caption{Probability for the hypothesis $\lambda_T < \lambda_T^0$ \label{figcumvary}}
\end{center}
\end{figure}

Eventually we are interested in the probability for the hypothesis that the true frequency of accidents is smaller than or equal to the value predicted by PRA. Using these values, we can now obtain the analogue of Table \ref{praright}. Below, we display the relevant values of $\cumprob(\lambda_T^0 | \nlist)$ at the same points as in Table \ref{praright}. This is the probability that the true current frequency of accidents is smaller than the values given by the PRAs considered below.
\begin{table}[!h]
 \resizebox{\textwidth}{!}{
\begin{tabular}{ccccc}
\toprule
{\bf Reactor}& \multicolumn{2}{c}{\bf Probability PRA CDF is right}  & \multicolumn{2}{c}{\bf Probability PRA LRF is right}  \\ \cmidrule(r){2-3}
\cmidrule(l){4-5}
&{\bf Fukushima separate} & {\bf Fukushima together} & {\bf Fukushima separate} & {\bf Fukushima together} \\ \midrule
 {Kudankulam}&$7 \times 10^{-24}$&$2 \times 10^{-17}$&$ 3 \times 10^{-22}$&$5 \times 10^{-14}$ \\
 {EPR}& $3 \times 10^{-16}$ & $1  \times 10^{-11}$ & $5 \times 10^{-17}$ & $2 \times 10^{-10}$ \\  
{AP1000}&$1 \times 10^{-17}$&$1 \times 10^{-12}$&$1 \times 10^{-17}$& $6 \times 10^{-11}$\\ \bottomrule
\end{tabular}}
\caption{Comparing PRA results with Bayesian estimates from historical observations after allowing exponential increase in safety standards \label{prarightsophisticated}}
\end{table}
We have also plotted this probability as a function of $\lambda_T^0$ in Figure \ref{figcumvary}.

It is sometimes believe that since safety standards have been improving over time, the results of PRA may be valid for newer reactors even if they are inconsistent with the empirical data for older reactors. The calculations above rule out this possibility and reinforce our conclusion that the results of PRA cannot be reconciled with empirical data.

\subsection{Regional Variations \label{secregional}}
It is clear that the safety of nuclear reactors may vary across regions. However, just as in our analysis of possible improvements in safety over time, we now show that no reasonable variation in safety across regions can lend any confidence to the results of PRA-calculations. 

In the case of regional variations, it would be inappropriate to proceed along the lines of \eqref{varylambda} since there is no reason to expect that safety will vary monotonically along any spatial parameterization of nuclear reactors. 

However, the central point is that for any particular country or region,  the experience of nuclear accidents in the rest of the world provides a reasonable {\em prior} estimate of the frequency of accidents in that region. If we have reason to believe, from the record of smaller incidents or from some other information, that safety in that region is better or worse than other parts of the world, we can account for this factor as well in our prior distribution. This prior distribution can then be corrected using empirical data from the region itself to obtain a final posterior distribution for the frequency of accidents. 

We now describe, more precisely,  how this procedure can be implemented. Consider a particular region, $R$,  which may be a country or a group of countries.  In this region, we assume that the probability of an accident in the interval $[t, t+dt]$ is given by \eqref{centralassumption}.
\be
d p = \lambda_R d t
\ee
In this subsection, we do not consider additional variations of accident frequencies with time,  to avoid complicating the analysis. In addition, we have the complement of the region (the rest of the world), $\widetilde{R}$, and we assume that in $\widetilde{R}$, the probability of an accident in an interval of length $d t$ is given by
\be
d p = \widetilde{\lambda}_{R} d t
\ee

Assuming a flat prior distribution for $\widetilde{\lambda}_R$, we now construct a posterior probability distribution for this parameter. Assuming that $n_{\rcomp}$ accidents have been observed in the rest of the world in a total operating time $T_{\rcomp}$, we use the techniques of section \ref{frequencypra} to construct the posterior probability distribution for $\lcomp$, which is given by
\be
\probdens(\lcomp | \ncomp) = {1 \over \Gamma(\ncomp + 1) } \Tcomp 
(\lcomp \Tcomp)^{\ncomp} e^{-\lcomp \Tcomp}.
\ee

Now, we then additionally input the assumption that the prior distribution for $\lambda_R$ is the same as the posterior distribution for $\lcomp$, except for a constant of proportionality $\kappa$ between them.  The factor $\kappa$ indicates our prior belief that nuclear safety in region $R$ is better or worse than that in other parts of the world. 

For example, we could obtain an estimate for $\kappa$ using a record of less severe incidents (not necessarily core-damage or large-release) that have occurred  in $R$ and $\widetilde{R}$
\be
\kappa = {I_{R} \over \widetilde{I}_{R}},
\ee
where $I_{R}, \widetilde{I}_{R}$ are the number of incidents recorded in $R$ and $\widetilde{R}$ respectively. It is, of course, necessary to have a precise criterion to count the incidents above, and we give one example below.  Since incidents of lower severity are fairly frequent, therefore  a purely frequentist analysis is sufficient to obtain an estimate for $\kappa$. 

After fixing $\kappa$ using this technique, or some other, we then take the prior distribution for $\lambda_R$ to be
\be
\label{priorR}
\probdens(\lambda_R) = \left. {1 \over \kappa} \probdens(\lcomp | \ncomp) \right|_{\lcomp = {\lambda_R \over \kappa}} = {1 \over \kappa^{\ncomp + 1} \Gamma(\ncomp + 1) } \Tcomp 
(\lambda_R \Tcomp)^{\ncomp} e^{-{\lambda_R \Tcomp \over \kappa}}
\ee

Now, denoting the number of accidents observed inside region $R$ by $n_R$, in a total operating time $T_R$, we can construct a posterior probability distribution for $\lambda_R$ using the same techniques and the prior in \eqref{priorR}. This leads to
\be
\probdens(\lambda_R | n_R) = {T_R + {\Tcomp \over \kappa} \over \Gamma(n_R + \ncomp + 1)} ((T_R + {\Tcomp \over \kappa}) \lambda_R)^{\ncomp + n_R} e^{-\lambda_R (T_R + {\Tcomp \over \kappa})}
\ee
The probability for the hypothesis that $\lambda_R < \lambda_0$ is given by
\be
\cumprob_R(\lambda_0, n_R) = {\Gamma(1 + n_{R} + \ncomp, (T_{R} + {\Tcomp \over \kappa}) \lambda_0) \over \Gamma(1 +
  n_{R} + \ncomp)}.
 \ee
So, we see that in this model, we obtain a simple result. The rule is that to account for variations in safety across regions, we simply count the total accidents $\ncomp + n_R$ as having occurred in a time
$T_R + {\Tcomp \over \kappa}$. In the situation where we take $\kappa = 1$, we recover the results of section \ref{frequencypra} since $T_R + \Tcomp$ just becomes the total operating experience $T_{\text{obs}}$. 

Now the key point is as follows. If we recall the expansion of the incomplete gamma function shown in \eqref{incompgammafunc} we find that for small $\lambda_0 (T_R + {\Tcomp \over \kappa})$ the expressions above are well approximated by
\be
\cumprob_R(\lambda_0, n_R) \approx {\left(\lambda_0 (T_R + {\Tcomp \over \kappa}) \right)^{n_R + \ncomp + 1}  \over \Gamma(n_R + \ncomp + 2)}
\ee
If we now take $\lambda_0$ to be one of the values given in Table \ref{tablepra} we see that the expression above necessarily evaluates to an extremely small number. 

To take a numerical example, let us take the region under consideration to be France. At the end of 2012, France had accumulated about $T_R = 1874$ reactor-years of operating experience, without a single large-release event. (See Table 4 in \cite{npriaea2013}.) In the same period the rest of the world had accumulated $\Tcomp = T_{\text{obs}} - T_R = 13373$ reactor-years of operating experience.  We also take $\kappa = 0.5$, which suggests a prior belief that French reactors are twice as safe as the world-average. We emphasize that this value of $\kappa$ is being taken here just as an example, and not to suggest that this is really the case.

With these parameters, given that there have been five large release events in the rest of the world, and taking the PRA frequency estimate for the French EPR reactor given above, we see that 
\be
\cumprob_R(7.69 \times 10^{-8}, 5)  = 1.6 \times 10^{-19}.
\ee
Therefore, given the existing empirical experience in the rest of the world, even with an assumption that French reactors are considerably safer, the probability that the EPR reactor genuinely has a true frequency of accidents as small as the predictions of PRA is absurdly low.

It is clear that the models in this section can be extended further to account for more detailed variations.   However, the analysis of this section shows that the severe contradiction between the results of PRA and empirical data cannot be resolved in any such manner. 

\section{On the Indian Experience \label{secindia}}
We now turn to the Indian experience. The correct procedure to analyze this case would be to use the rest of the world's experience as a prior distribution as was done in section \ref{secregional}. Needless to say, with any reasonable choice of $\kappa$, as defined there, we would end up with the conclusion that $\epsilon \ll 1$ for India. However, in this short subsection, we want to briefly take a separate approach to make an elementary statistical point. Even if one assumes that the existing record of nuclear accidents has absolutely no bearing on the Indian situation,  India's operating experience of  $T_{\text{ind}} = 394$ reactor years  \cite{npcilmain} is too low to provide any statistical confidence in the safety of the Indian nuclear programme. 

The significance of this observation pertains to common claims made by the Indian nuclear establishment that Indian reactors are safer since India has not witnessed a major accident in this time period. The point of this section is to point out that such claims are statistically fallacious. While we have phrased our discussion in India's context, it is worth noting that several countries have accumulated similar levels of operating experience. To take a few examples, by the end of 2012, Belgium had accumulated $254$ reactor-years, China had accumulated $141$ reactor-years, Canada had accumulated $634$ reactor-years, the republic of Korea had accumulated $404$. Although none of these countries have seen a major accident, the conclusions of this section apply to all of them: their operating experience is insufficient to suggest that safety levels in these countries are significantly different from the world-average.

In mathematical terms, in this section our objective is somewhat different from the rest of the paper. Here, we are not trying to estimate $\epsilon$ (from \eqref{centralclaim}) and show that it is very small, but rather we would like to show that even if one discards the reasonable prior assumptions made in section \ref{secregional}, there is no reason to suppose that $\epsilon \sim 1$ for these countries. As explained above, the numerical figures that we take below are specific to India, but the reader can easily modify this calculation to the other countries mentioned above.  
 
\begin{figure}[!h]
\begin{center}
\includegraphics[width=0.6\textwidth]{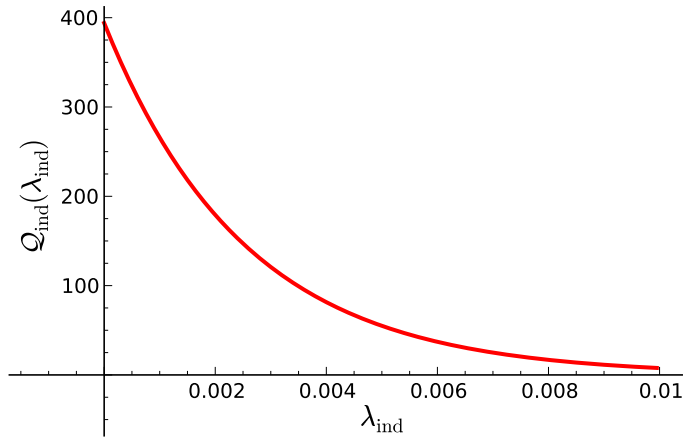}
\caption{Posterior probability distribution for 
  $\lambda_{\text{ind}}$ \label{pind}}
\end{center}
\end{figure}

There have been several minor but no major accidents in India's operating history. 
We again start with a
flat prior distribution for the mean frequency of accidents at Indian reactors, which we denote by $\lambda_{\text{ind}}$ to distinguish it from the global frequency. Then the posterior distribution 
for $\lambda_{\text{ind}}$ is given by
\be
{\mathcal{Q}}_{\text{ind}} (\lambda_{\text{ind}} ) = T_{\text{ind}} e^{-\lambda_{\text{ind}}
  T_{\text{ind}}},
\ee
Note we can obtain this by setting $n_{\text{obs}} \rightarrow 0$ and $T_{\text{obs}} \rightarrow T_{\text{ind}}$ in \eqref{posterior}.
This curve is plotted in figure \ref{pind}.

This figure tapers off quite gently, so India's current operating
experience cannot tell us much about the frequency of accidents in India, especially if the true frequency is around $\nu^{\text{cd}}_{\text{obs}}$ in \eqref{observedfreq}.

\begin{figure}[!h]
\begin{center}
\includegraphics[width=0.6\textwidth]{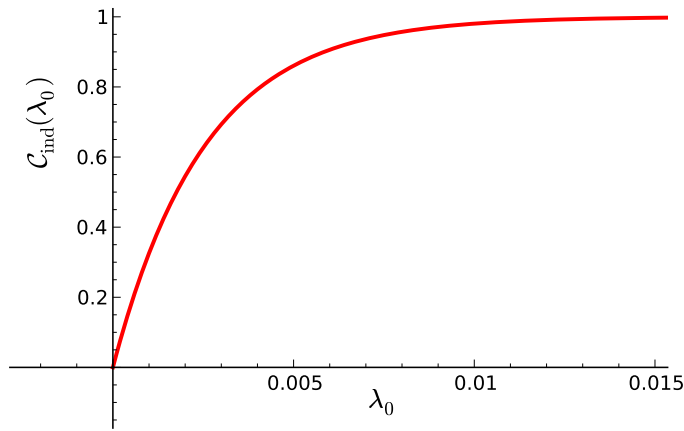}
\caption{Probability for the hypothesis $\lambda_{\text{ind}} < \lambda_0$ \label{cumprobind}}
\end{center}
\end{figure}

To make this sharper, it is useful to
look at the probability for the hypothesis that
$\lambda_{\text{ind}} < \lambda_0$. This probability is given 
 by the function
\be
{\cumprob}_{\text{ind}} (\lambda_0) = \int_0^{\lambda_0} {\cal
  Q}_{\text{ind}} (\lambda) \, d \lambda = 1 - e^{-T_{\text{ind}} \lambda_0}.
\ee
This curve is shown in figure \ref{cumprobind}.

Some relevant numerical figures are
\be
\begin{split}
&{\cumprob}_{\text{ind}}(2.67 \times 10^{-4}) \approx 0.10, \\
&{\cumprob}_{\text{ind}}(1.0 \times 10^{-3}) \approx 0.33.
\end{split}
\ee
In words, this indicates that it is only with a confidence of about 10\%
  that we can state that the true frequency of accidents is smaller than
 once in 3700 reactor-years. 
  And it is only with a confidence of about 33\% that we can conclude that
  the true frequency of accidents in India is smaller than once per thousand reactor-years.

These results are very simple to understand intuitively. They simply
reflect the fact that even if the expected frequency of accidents in India is once every thousand reactor-years, there is still an excellent chance that one could get through $394$ reactor-years
without an accident. Conversely, the absence of an accident in this time period
is not particularly informative.

It is quite common to find claims like the one made by the Indian Nuclear Society its advertisement for a 2012 conference. ``Having achieved safe and reliable operation of about 360 reactor- years \ldots   the Indian nuclear programme has demonstrated a high level of maturity. The safety track record of Indian nuclear power plants has been impeccable'' \cite{insac2012}.
These claims are accepted at the highest levels of government. In 2011, the Prime Minister stated that  ``The safety track record of our nuclear power plants over the past 335 reactor-years of operation has been impeccable'' \cite{pmdaespeech}.
Our analysis shows that it is erroneous to draw these complacent conclusions. 
 India's operating experience is too limited to 
provide statistically reliable long term estimates about the efficacy or otherwise of safety practices in the nuclear sector. 

In passing, we should point out that, within the nuclear industry, it appears
to be a rather common statistical error to extrapolate from limited experience
to rather strong claims of safety. 
For example, the Rasmussen report also stated that ``It is significant that in some 200 reactor-years of commercial operation of reactors of the type considered in the report there have been no fuel melting accidents.'' (See p. 6 of  \cite{rasmussen}.) However, this experience had absolutely no significance for the conclusion drawn by the report, which was that the core-damage frequency of reactors was once in 20,000 years (p. 8).

\section{The Debate on Nuclear Liability \label{secliability}}

It is clear that the results of PRA are untenable in the light of empirical data. In this section we provide evidence, using the debate on nuclear liability, that the industry's actions (as opposed to its
public statements) suggest that it has independently reached this conclusion.

Soon after the nuclear deal, multinational nuclear suppliers lobbied
the government to pass a law that would indemnify them in 
the event of an accident. Under this pressure, the government 
passed a liability law in 2010
that was almost identical 
to the annex of a U.S. sponsored
international convention on the subject called the Convention on Supplementary Compensation \cite{suvratramanaliab,suvrat_raju_strange_2011}. However, as a result of various pulls-and-pushes in the legislative process, nuclear 
suppliers are largely protected but not completely indemnified by the Indian law. While victims cannot sue the supplier, the law allows the NPCIL, which has the primary
responsibility of compensating victims, to recover some of this compensation 
from the supplier 
for a disaster caused
by substandard equipment. The refusal of suppliers to accept even this marginal 
liability has presented a significant obstacle for contracts for new reactors. (See for example, the recent statement by the CEO of General Electric \cite{gerefusal2015}.)

It is significant that although one of primary advertised benefits of the Indo-US nuclear deal was that India would be able to purchase light water reactors from new suppliers, the deadlock on liability has prevented this entirely. The first new contract for reactors after the nuclear deal was signed in 2014 with the Russian public sector company Rosatom. However, this will just extend the Kudankulam nuclear complex which was covered by an inter-governmental agreement even before the nuclear deal.  Negotiations with suppliers from markets that have opened up after the nuclear deal, including companies from France and the United States, have so-far failed to yield a single new contract due to the conflict on liability.

However, this leads directly to the following question: if the chance of a nuclear accident is 
indeed as remote as the industry claims, then why are nuclear reactor
manufacturers unwilling to accept liability for an accident? 

One of the ostensible reasons given by suppliers is that forcing them to accept liability would cause the cost of power to go up \cite{russiandeputypm}. 

 To examine the veracity of this claim, consider the  expected cost of insurance for suppliers if the results of PRA are taken at face value by the industry and actuaries \cite{suvratramanahindu2012}.

The Indian liability law currently caps the total available compensation for victims at ``the rupee equivalent of three hundred million Special Drawing Rights''(SDRs). (See clause 6 of the text of the law  \cite{gazetteliability}.) This includes compensation from the operator, and also the central government, and victims are not legally entitled to any further compensation.  Here we consider the worst case scenario for the supplier, where it becomes liable for this entire amount. 

Although the rupee to SDR exchange rate fluctuates, this maximum liability is approximately 
$\ell_{\text{cap}} = \text{Rs.}~2,500~\text{crores}=\text{Rs.}~2.5
\times 10^{10}$ in rupee terms. 

Taking, for example, the claimed frequency of core-damage accidents at the Kudankulam reactors, which we denote by $\mu_{\text{kk}} = 10^{-7} \ry$, then a simple order
of magnitude estimate for the cost of insurance for this amount is 
\be
i_p = \mu_{\text{kk}} \times {\ell}_{\text{cap}} = \text{Rs.}~
2,500.
\ee
 A reactor with a capacity of 1000 MW, operating at a 80\% load-factor, should produce 
$E = 0.8 \times 10^6 \times 365 \times 24~ \text{kWh} \approx 7 \times 10^9~ \text{kWh}$ 
of electricity each year. So, the cost of insurance above should lead to an
increase in the cost of electricity by 
\[
\delta p = {i_p \over E} = 3.6 \times 10^{-7}~ \text{Rs.} /\text{kWh}!
\]
This absurdly small number indicates that something is amiss with the industry's claim that liability
will lead to price increases.

In fact two factors of about about $10^3$ and $10^4$, which are missing
in the calculation above, are required to make sense of the industry's reluctance. The first is that nuclear accidents could lead to damage that is a thousand times more than the cap on liability. 
For example, some estimates of the economic damage at Fukushima are as high as  $\ell_{\text{real}} \approx \text{USD}~200~\text{billion} \approx ~\text{Rs.}~12~\text{lakh~crores}$. (See Table 4 of \cite{jcerfukushima}.)

In principle, a future Indian government could ignore the liability cap and insist
on recovering larger costs from the supplier. 
Even this would not lead to a prohibitive cost of insurance if reactors
were genuinely as safe as manufacturers' claim.
The other crucial missing factor comes from our result above: accidents 
affecting the public are likely to happen at a rate that is closer to $\nu^{\text{lr}}_{\text{obs}}$. 
Extending our simple linear model with these realistic estimates of damage
and risk leads to the following cost of insurance per unit energy produced
\[
\delta p_{\text{true}} = {\nu^{\text{lr}}_{\text{obs}} \times \ell_{\text{real}} \over E} = 0.56~\text{Rs.}/\text{Kwh}.
\]

This is now a significant fraction (roughly 10\%) of the cost of electricity. However, at this point, corrections to our linear model for the insurance premium become significant. 
For example, since the total amount involved, $\ell_{\text{real}}$, is very high, and the  
expected rate $\nu^{\text{lr}}_{\text{obs}}$ is non-negligible,  financial institutions would evidently
be unwilling to underwrite this risk without additional incentives
in the form of a significantly higher cost of insurance.   This helps explain why suppliers insist on legislative indemnity, rather than simply arranging
for the appropriate financial cover.

What the debate on liability shows is that the nuclear industry --- both in the private and the public sector --- has itself taken note of the empirical
rates of accidents, and it is unwilling to take the predictions of its PRAs seriously when its economic
interests are at stake.

\section{Conclusions \label{secconclude}}
In this paper, by means of some simple Bayesian calculations, we have come
to the following conclusion:  the historical record contradicts the predictions of probabilistic risk assessment and suggests a significantly higher risk of nuclear accidents. The contradiction between these predictions and data can be quantified in terms of a probability for the hypothesis that true frequencies of accidents are as small as those predicted. This probability can be quantified in various models, and is found to be extraordinarily small in a model-independent fashion, and independent of changes in our detailed assumptions. In particular, in section \ref{secrobust} we showed how, even in models where safety standards improve exponentially with operating experience or where reactors in a given region are assumed to be considerably safer than reactors elsewhere, the conclusions above hold at very high confidence levels. 

Second, we also specifically discussed the case of India. Although this is strictly speaking, a subset of the study of worldwide accident frequencies above, we analyzed it separately to show that India's current India's current  experience with nuclear reactor operation is far from sufficient to draw any strong conclusions about future reactor safety. The same conclusions hold for other countries with similar levels of operating experience, such as Canada, China, Belgium and Korea.

Therefore it is clear that the methodology and practice of PRA needs to be revised significantly. 

In fact, as we showed in section \ref{secliability}, it is clear from the debate on nuclear liability that the nuclear industry already recognizes that the results of PRA are numerically unreliable. Nevertheless, within the technical community, the use of PRA is sometimes justified as a useful tool for safety analysis. For example, the authors of \cite{sornette2013exploring} explained that even though PRA is ``not thought to represent the true risks'' it remains useful as a ``platform for technical exchanges on safety matters between regulators and the industry.'' However, this begs the question: why has PRA failed so badly in achieving the purpose that it was designed for. Indeed, the Rasmussen report \cite{rasmussen} started by explaining that ``the objective of the study was to make a realistic estimate of [the] risks'' that would be ``involved in potential accidents.'' 

Apart from the theoretical problems mentioned in the introduction, it appears likely that the nuclear industry benefits from the disingenuous suggestion that it can, in fact, accurately predict the frequency of accidents. Although insiders recognize that this is not the case, it is clear that the detailed computer simulations that support a supposedly-scientific calculation of low accident-frequencies computed to several decimal places are useful in public debates. While there have been other critiques of the mismatch between the results of PRA and empirical data, we believe that this study is significant because it emphasizes the extraordinarily high level of confidence with which it is possible to rule out the results of PRA.

Indeed, this would hardly be the only situation in which the nuclear industry has attempted to use the authority of science to dismiss safety concerns. For example, in an attempt to dismiss the history of Chernobyl, the World Nuclear Association declared \cite{wnasafetyarchive2011} in January 2011 that ``In the light of better understanding of the {\em physics and chemistry} of material in a reactor core  \ldots it became evident that even a severe core melt coupled with breach of containment could not in fact create a major radiological disaster from any Western reactor design'' (emphasis added). After attempting to initially defend this claim for a few days after Fukushima by claiming that ``clearly there was no major release from the reactors''\cite{wnasafetyarchive2011_afterfuk} and only from the ``fuel pools'', the Association had to reluctantly concede that its claim, seemingly based on rigorous material science, ``did not apply to all'' \cite{wnasafetyarchive2011_concession} Western reactor designs. 

It is interesting to note that a similar dynamic operates in other industries as well. For example,  in the aviation industry (which inspired the use of
PRA for nuclear reactors), as part of the certification process for its new 787 ``Dreamliner'' aircraft, Boeing estimated that its lithium-ion batteries would vent smoke ``once in every 10 million flight hours.'' In fact this event occurred twice in 52,000 flight hours leading to the grounding of the entire
fleet for inspection \cite{dreamlinerinterim}
 
To explore the implications of our conclusions, we return to the case of India where the government is planning a large nuclear expansion. 
It has announced plans to commence construction on 8 heavy water reactors, with a capacity of 5600 MW in the ``12\textsuperscript{th} Plan'' period (2012--17), and complete work on a separate 2800 MW of installed capacity. In addition, it is also planning to import 8 reactors with a total capacity of 10,500 MW \cite{narayanasamy_setting_2012-1}. 
Every reactor site has seen
vigorous local protest movements that have raised issue of land and livelihood but also questions about nuclear safety. 

Therefore it is imperative to have a frank conversation on nuclear safety, involving not just the technical community but a far broader cross-section of society. Our results show that 
such a debate should start with the acceptance that the ambitious
claims about nuclear safety made on the basis of probabilistic risk assessment
have been conclusively falsified by the empirical data.

\paragraph*{Acknowledgments}
 This article is based on a detailed exchange of letters with Mr. Nalinish Nagaich of the NPCIL. I am grateful to M. V. Ramana for comments on a draft of this manuscript. 

\bibliographystyle{ieeetrurl}
\bibliography{refs}

\end{document}